\documentclass[twocolumn,showpacs,prl,preprintnumbers,amsmath,amssymb]{revtex4}

\usepackage{dcolumn}% Align table columns on decimal point
\usepackage{bm}% bold math
\usepackage{graphicx}
\usepackage{amssymb}

\begin{document}

\title{Hyper-Raman scattering from vitreous boron oxide:\\
coherent enhancement of the boson peak}

\author{G. Simon, B. Hehlen, E. Courtens, E. Longueteau, and R. Vacher}

\affiliation{ Laboratoire des Collo\"{\i}des, Verres et Nanomat\'eriaux (LCVN), UMR 5587 CNRS, University of 
Montpellier II, F-34095 Montpellier, France}

\date{\today}

\begin{abstract}
Hyper-Raman scattering spectra of vitreous B$_2$O$_3$ are reported and compared to
Raman scattering results.
The main features are indexed in terms of vibrations of structural units.
Particular attention is given to the low frequency boson peak which is
shown to relate to out-of-plane librations of B$_3$O$_3$ boroxol rings and BO$_3$ triangles.
Its hyper-Raman strength is comparable to that of cooperative polar modes.
It points to a sizeable coherent enhancement of the
hyper-Raman signal compared to the Raman one.
This is explained by the symmetry of the structural units.
\end{abstract}


\maketitle

Most glasses exhibit at low frequencies $\omega$ a broad spectral component observed both
in Raman scattering (RS) \cite{Shu71,Bro96} and inelastic neutron-scattering (INS) \cite{Buc86,Eng98}.
The reduced density of vibrational states, $Z(\omega )/\omega ^2$, usually shows in that region
a considerable excess of harmonic modes over the Debye level of sound waves \cite{Phi81}.
This is called the boson peak (BP).
Although the nature of BP-vibrations remains much debated, it is generally
accepted that it will be key to a real understanding of the glassy state.
Lower frequencies being generally associated with larger objects, the BP should provide
indirect structural information at distances of one to a few nm.  
The structure of glasses at those scales is of considerable fundamental and practical interest, 
but it is generally not accessible to standard structural analysis tools.
The nanoscale also corresponds to the size of crystal nuclei, so that 
phenomena at that scale are relevant to the glass formation itself.
The slow progress in understanding the BP relates to the lack of experimental
techniques to address the structural issues.
For this reason, spectroscopies able to complement the information obtained with RS and INS are particularly
valuable.
Recently, hyper-Raman scattering (HRS) was found to be a powerful 
tool for the investigation of low $\omega$ excitations in vitreous silica, $v$-SiO$_2$ \cite{Heh00}.
These experiments confirmed previous measurements \cite{Buc86} and simulations 
\cite{Gui97}, indicating a large contribution of rigid SiO$_4$ tetrahedra librations in 
 the BP of $v$-SiO$_2$.
This single example already demonstrates that the microscopic origin of the BP 
follows from the specific molecular structure of the glass.

Boron oxide is the second most important glass former after $v$-SiO$_2$.
Boron forms covalent bonds with oxygen, and $v$-B$_2$O$_3$ is a network of oxygen-sharing BO$_3$ triangles.
Three such triangles can associate to produce a planar supermolecular unit, the boroxol ring B$_3$O$_3$.
The fraction of boron belonging to boroxol rings is difficult to ascertain from diffraction results alone \cite{Han94}.
However, NMR \cite{Zwa05} as well as recent simulations of spectra \cite{Uma05} have now confirmed that approximately
3 out of 4 borons belong to boroxol rings.
Hence, the ratio in the number of rings and independent triangles is about 1/1.
In this Letter, we report the first HRS results on $v$-B$_2$O$_3$.
We show that the HRS BP can be associated with out-of-plane librational motions of rigid rings and triangles.
In addition, the HRS intensity of the BP is exceptionally high, of the same order as
that of collective longitudinal optic (LO) and transverse optic (TO) polar excitations.
This suggests that the coherence of the BP vibrations is quite extended in space.
This coherence contributes much more effectively to HRS than to RS spectra for symmetry reasons that are explained below.

The HRS measurements were performed on a new spectrometer optimized for the study of liquids and disordered solids.
It combines high luminosity with a favorable resolution and contrast, allowing to investigate low-frequency modes. 
The incident infrared radiation at $\lambda$=1064 nm is delivered by a $Q$-switched Nd:YAG laser, producing
$\sim$23 ns-long pulses of up to 70 kW peak power at 2.5 kHz repetition rate.
These are focused with a $f$=3 cm lens, down to a beam waist in the sample of $\sim 10$ $\mu$m radius.
The scattered light is collected with a $f$/1.8 aperture.
The spectrum is analysed with a single grating monochromator and detected with a N$_2$-cooled CCD camera.  
Two gratings are available, with 600 or 1800 groves/mm.
Using an entrance slit of 100$~\mu$m, this leads to instrumental full widths at half 
maximum of $\sim 6$ or $\sim 2$~cm$^{-1}$, respectively.
The RS spectra shown for comparison were obtained with 514.5 nm excitation and
analyzed with a Jobin-Yvon T64000 triple monochromator, operated in the macromode.
The $v$-B$_2$O$_3$ sample was prepared from isotopically pure (99.6\%) $^{11}$B$_2$O$_3$ containing $\sim 1$ wt \% moisture.
The material was heated to 1100 $^\circ$C in a platinum crucible, quenched on a heat conducting plate,
and annealed at 571 $^\circ$C for 200 hours.
The sample was cut and dry-polished.
The water content (0.8 mole \%) was subsequently measured by IR-transmission of a thin slice as described in \cite{Ram97}.
The same sample, sealed in an optical silica cell containing silica gel, was used for RS and  HRS measurements in 90$^\circ$-scattering.

Figure 1 shows polarized (VV) and depolarized (VH) RS and HRS susceptibilities $\chi '' = I(\omega)/(n+1)$.
$I(\omega)$ is the spectral intensity and $n$ the Bose factor.
These HRS spectra were obtained using the 600 groves/mm grating.
It is evident that there are many modes and that the RS and HRS spectra are remarquably dissimilar.
\begin{figure}
\includegraphics[width=8.5cm]{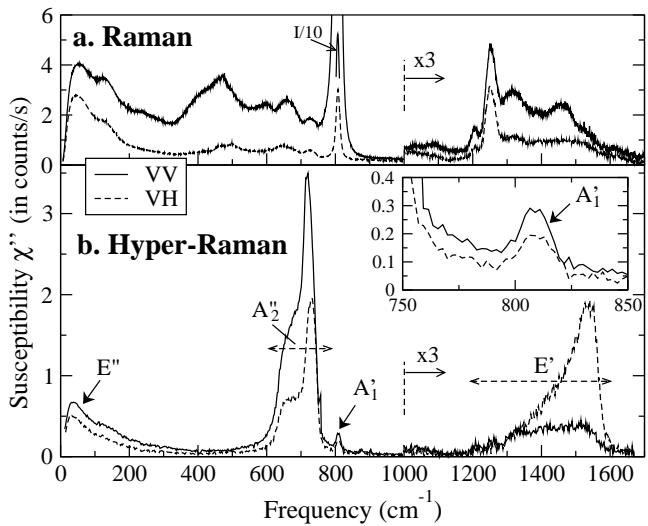}
\caption{Raman (a) and hyper-Raman (b) susceptibility spectra of $v$-B$_2$O$_3$ at room temperature.
In VV, the main RS feature at 808 cm$^{-1}$ is off scale. Its peak is drawn divided by 10.
Above 1000 cm$^{-1}$, the vertical scale is expanded by 3 for all spectra.
The inset shows on an enlarged scale the HRS-spectra of the 808 cm$^{-1}$ mode.}
\end{figure}
However, one should note that the RS presentation zooms on small features, with the largest peak way off scale, while the HRS 
ordinate is adjusted to the dominant features, with small peaks mostly buried in the background.

The large number of modes is related to the simultaneous presence of BO$_3$ triangles and B$_3$O$_3$
rings \cite{Gal80}, in nearly 1/1 ratio.
We remark that the sum of one triangle plus one ring just amounts to two formula units 2(B$_2$O$_3$).
In a simplified model, triangles and rings can thus be taken as the two building blocks.
Both have rhombohedral D$_{3h}$-symmetry.
Their vibrations are decomposed into irreducible representations:
\begin{eqnarray}
{\rm [1,2]\; A'_1 (R,HR) + [1,2]\; A'_2(HR) + [2,2]\; A''_2(IR,HR) +}\nonumber\\
{\rm [3,4]\; E'(IR,R,HR) + [1,2]\; E''(R,HR) }\; \; . \; \; \; \; \; \; \; \; \;
\end{eqnarray}
The figures within square brackets show the number of independent representations for triangles and
rings, respectively, while the symbols in parentheses indicate activity in infrared (IR), Raman (R), or hyper-Raman (HR).
The notation for the representations is that of \cite{Wil55}.
Our purpose is not to develop a detailed calculation of the spectra.
It is rather to gain qualitative understanding in the nature of the main HRS modes and in the relative RS and HRS strengths.
To this effect the above model, admitedly rather simple, should be sufficiently realistic.
For each type of unit, $\rm A'_1$ are radial breathing motions, $\rm A'_2$ are in-plane librations,
one $\rm A''_2$ representation is an out-of-plane translation while the other is a polar internal vibration,
similarly one $\rm E'$ representation corresponds to in-plane translations while the others are polar internal vibrations,
and finally $\rm E''$ are out-of-plane librations \cite{Wil55}.

In this molecular picture, light scattering results from the fluctuations of dipoles induced
on all structural units by the applied optical field {\bf E} $ \propto e^{-i \omega t}$.
For unit $i$, the induced dipole can be written ${\bf p}^i = 
{\bm  \alpha}^{i} \cdot {\bf E'} + {\bm \beta}^{i} : {\bf { E' E'}} + ...$, where 
${\bm  \alpha}^{i}$ and ${\bm \beta}^{i}$ are the first and second-order polarizability tensors, 
respectively, and ${\bf E'}$ is the local field.
In the molecular reference frame \cite{foot}, the form of ${\bm  \alpha}$ and ${\bm \beta}$ is well known.
For a symmetric top, ${\bm  \alpha}$ is diagonal with $\alpha_{xx}=\alpha_{yy} \neq \alpha_{zz}$.
For ${\bm \beta}$, we remark that both $\omega$ and $2\omega$ are far from any material resonance in our experiment.
In that case, ${\bm \beta}$ can be approximated as fully symmetric in all permutations of its three indices \cite{Den87}.
For D$_{3h}$-symmetry, there remains just $\beta_{xxx}=-\beta_{xyy}$ with all other index combinations zero.
For unit $i$, ${\bm  \alpha}^{i}$ and ${\bm \beta}^{i}$ are just rotated from the molecular frame to the fixed laboratory frame.
It is the modulation of ${\bm  \alpha}$ and ${\bm \beta}$ by the normal modes, of amplitude
$W^\zeta \propto e^{\pm i \omega_\zeta t}$ for mode $\zeta$, which leads to the molecular 
Raman and hyper-Raman polarizability tensors, ${\bm \alpha}^\zeta$ and ${\bm \beta}^\zeta$, respectively.
These have been tabulated for all molecular symmetries, {\em e.g.} in \cite{Cyv65}.
In the laboratory frame, each rotated molecule $i$ has then a fluctuating dipole at $\omega_\zeta$:
$$\delta{\bf p}^{\zeta,i} = {\bm \alpha}^{\zeta,i} \cdot {\bf E'} + {\bm \beta}^{\zeta,i} : {\bf { E' E'}} + ... \; \;. \eqno{(2)}$$
The local field corrections are essential for quantitative
results \cite{Jan02} but they should not affect our qualitative symmetry considerations.

Structural units that vibrate independently from each other, {\em i.e.} with random phases, scatter {\em incoherently}. 
The intensity is then just the sum of the individual intensities, $\propto \sum_{i} | \delta{\bf p}^{\zeta,i} | ^2$.
This is expected for all non-polar internal vibration modes in random media \cite{Cyv65}.
On the other hand, if several structural units vibrate with a fixed phase relationship, there is
{\em coherent} scattering, and the intensity is $\propto | \sum_{i} \delta{\bf p}^{\zeta,i} | ^2$.
This occurs in two important cases:
i) for {\em polar} vibrations, which are the only ones compatible with a collective modulation of the average
D$_{\infty h}$ glass symmetry \cite{Den87};
ii) for external modes, in particular librations for which nearby units move together, leading to overall coherence.

The narrow peak at 808 cm$^{-1}$ reaches over 50 in Fig. 1a, entirely dominating the VV-spectrum.
This component has been indexed either as the $\rm A'_1$ oxygen breathing of the boroxol rings \cite{Win82}, or
as a correlated symmetric stretch motion of the O-atoms over the network \cite{Mar81}.
The modulation of the boroxol $\bm \alpha$ by breathing only gives diagonal components, with
$\alpha_{xx}^\zeta=\alpha_{yy}^\zeta \neq \alpha_{zz}^\zeta$ ($\zeta=\rm {A'_1}$).
If there would be coherent breathing of adjacent boroxols, one would expect for steric reasons that they would
vibrate in opposite phase, so that their contributions $\delta{\bf p}^{\zeta,i}$ would mostly cancel each other.
Incoherent breathing of very regular boroxol units is able to produce the observed narrow peak.
This is now supported by a first-principle analysis \cite{Uma05}.
The observed strong polarization just indicates that the isotropic part, of trace $(2 \alpha_{xx}^\zeta + \alpha_{zz}^\zeta)$,
is much larger than the anisotropy $|\alpha_{xx}^\zeta - \alpha_{zz}^\zeta|$.
From (1), this vibration is equally active in HRS.
In Fig. 1b, only a relatively small peak is observed at the corresponding frequency.
It is shown in more details in the inset.
The modulation of $\bm \beta$ by $\rm A'_1$ leads to $\beta_{xxx}^\zeta=-\beta_{xyy}^\zeta$,
with all other combinations vanishing \cite{Cyv65}.
In incoherent scattering, this produces a depolarization ratio $I_{\rm VH}/ I_{\rm VV} = 2/3$ \cite{Cyv65}, in 
excellent agreement with our observation.
There is no reason to believe that HRS would be particularly small for this mode compared to RS.
Rather, the comparative sizes of the 808 cm$^{-1}$ peaks in Figs. 1a and 1b is presumably a fair
indication for the relative scale of these two plots.
This is supported by the observation of many modes between 300 and 600 cm$^{-1}$ in RS (Fig. 1a), while corresponding features
are not apparent in HRS (Fig. 1b).
We remark that all RS-active modes in (1) are also HRS-active.
Since the weak RS modes are just not seen on the scale of our HRS spectra,
we conclude that strong spectral components in Fig. 1b must all be enhanced by {\em coherent} effects.

This is known for the groups around 700 and 1400 cm$^{-1}$ which are
IR-active \cite{Ten72,Gal80}, and thus HRS-allowed in the mean isotropic symmetry D$_{\infty h}$.
The isotropic HRS tensor $\bm \beta$ has components $\beta_{\rm VVV} = 3b$ and $\beta_{\rm HVV} =b$ \cite{Pod96}.
The coupling of the LO-motions with the internal Maxwell field produces the LO-TO splitting.
The tensor elements $b$ are distinct for TO and LO modes, $b_{\rm TO}$ and $b_{\rm LO}$ respectively, the 
latter being generally much enhanced by the electro-optic coupling.
In 90$^\circ$-scattering, one finds that $I_{\rm VV} \propto 9 b_{\rm TO}^2$ while $I_{\rm VH} 
\propto  {\scriptstyle \frac {1}{2}} b_{\rm TO}^2 + {\scriptstyle \frac {1}{2}} b_{\rm LO}^2 $ \cite{Den87}.
From (1), the IR-active modes have $\rm A''_2$ and $\rm E'$ symmetries.
The modes of $\rm A''_2$-symmetry correspond to opposite displacements of B and O along the $z$-axis.
This type of motion has a significantly lower restoring force than the in-plane displacements of the $\rm E'$-modes.
Hence, the group around $700$ cm$^{-1}$ is assigned to the $\rm A''_2$ internal modes.
This double peak structure with maxima located at $\sim 655$  and $717$ cm$^{-1}$ in VV-scattering and previously
observed in IR-transmission \cite{Ten72}, must be due to the separate activity of triangles and rings. 
The narrowest peak at $717$ cm$^{-1}$ in VV-scattering is assigned to the boroxols, 
as  confirmed by a first-principle calculation \cite{Pas05}. 
Regarding the VV and VH spectra, we find for this peak a TO-LO splitting of $\sim 15 \pm 2$ cm$^{-1}$. 
This is within the accuracy of the value $\sim 20$ cm$^{-1}$ reported from IR-reflectiviy \cite{Gal80}.  

The group centered around 1400 cm$^{-1}$ is due to modes $\rm E'$ active in IR, RS, and HRS.
The main TO motion produces the feature at 1260 cm$^{-1}$.
Like for the breathing mode, it is much better seen in RS than in HRS.
It is related to the boroxol rings \cite{Has92}.
The LO-TO splitting is $\sim 290$ cm$^{-1}$ \cite{Gal80}, and the corresponding LO peak is the strongest feature in the VH-HRS spectrum.
The entire group is rather rich, with at least six peaks in the RS spectra.
This agrees qualitatively with the large number of different modes of $\rm E'$ symmetry in (1).

We are now equipped to discuss the nature of the BP vibrations.
Figure 2 shows in more details the $T$-reduced HRS intensities, $I/(n+1)\omega$,
obtained with the 1800 groves/mm grating.
\begin{figure}
\includegraphics[width=8cm]{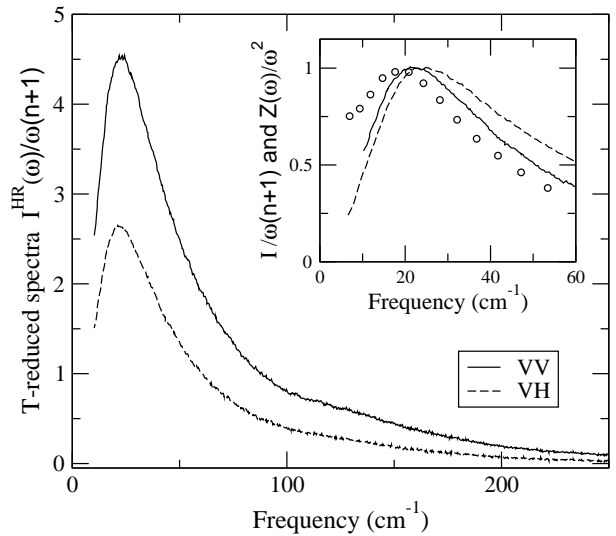}
\caption{High resolution 90$^\circ$-scattering HRS spectra of $v$-B$_2$O$_3$ at low $\omega$.
The inset compares the VH BP-spectra observed here in RS (dashed line) and HRS (solid line) 
with the total $Z(\omega )/\omega ^2$ obtained in INS at 330 K (dots) from \cite{Eng98}.}
\end{figure}
Since the BP is observed both in RS and HRS, its symmetry must be either $\rm A'_1$, $\rm E'$, or $\rm E''$.
The breathing modes $\rm A'_1$ occur at higher frequencies, such as 808~cm$^{-1}$ for the boroxols. 
The $\rm E'$ modes, as just shown, mostly belong to the high-frequency band around 1400 cm$^{-1}$.
However, they do include relative translations that can be of rather low frequency and which, in a
collective model, must couple to rigid rotations.
The modes of symmetry $\rm E''$ are precisely the out-of-plane rotations.
They include the librations of rigid triangles and rings, plus an internal out-of-plane deformation mode of 
rings that must be of higher frequency.
The rigid librations, coupled to relative translations, remain the only reasonable candidates for the BP.
That the librations of structural units produce the lowest frequency modes was already
noted for other glasses, {\em e.g.} $v$-SiO$_2$ \cite{Buc86,Heh00} or Se \cite{Ber94}.
The relative scattering strengths in Fig. 1 also suggest a {\em coherent} contribution that greatly enhances the
HRS-signal of these rigid librations.
This is further supported by the HRS depolarization ratio $I{\rm _{VH}}/ I{\rm _{VV}}$ which is $0.58$ at the peak
maximum in Fig. 2.
This is significantly smaller than the value 2/3 that applies to purely incoherent scattering \cite{Cyv65}.
\begin{figure}
\includegraphics[width=8.5cm]{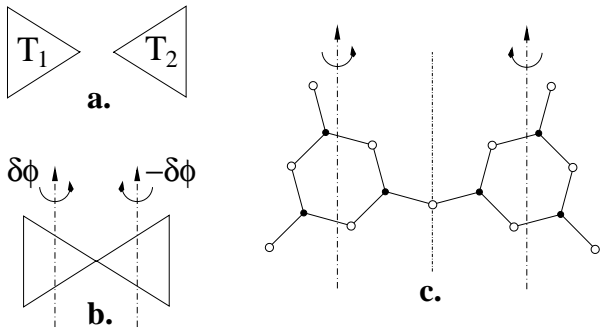}
\caption{a. The relative position of two D$_{3h}$ objects, T$_1$ and T$_2$, in a plane;
b. The joint libration of these rigid objects produced by a vertical displacement of the common apex;
c. The libration motion of two adjacent boroxols lying in a plane.}
\end{figure}

The coherent enhancement can be qualitatively understood from the symmetry properties of the polarizability
and hyper-polarizability tensors.
Equilateral triangles are the basic D$_{\rm 3 h}$-symmetry objects.
Fig. 3a shows two triangles T$_1$ and T$_2$, where T$_2$ is derived from T$_1$ by a 180$^\circ$-rotation.
It is easy to see that the $\bm \beta$-tensors of T$_1$ and T$_2$ are related by ${\bm \beta}^1 = - {\bm \beta}^2$.
However, the linear polarizabilities are identical, ${\bm \alpha}^1 = {\bm \alpha}^2$.
Now, suppose that these triangles share an apex that oscillates vertically, so that they librate by
angles $\pm \delta \phi$, as drawn in Fig. 3b.
The modulation of $\bm \beta^1$ by $+\delta \phi$ is equal in sign and magnitude to that of $\bm \beta^2$
by $-\delta \phi$, so that the HRS dipole fluctuations of both triangles add coherently.
On the contrary, the modulation of $\bm \alpha^1$ by $+\delta \phi$ is opposite to that 
of $\bm \alpha^2$ by $-\delta \phi$, so that these cancel and there remains no coherent RS-contribution.
This trend remains for out-of-plane librations of a connected set of triangles, also for 
non-planar arrangements: cancellations preferentially occur for $\bm \alpha$ rather than for $\bm \beta$.
Fig. 3c illustrates in a simple planar case how this can be extended to boroxol rings.
It shows two rings connected by a single B-O-B bridge.
The angle of the external B-O-B bond is $\sim 135 ^\circ$ \cite{Uma05}.
The relative rotation of the two boroxols by only $\sim 15 ^\circ$ is sufficient to produce a $\bm \beta^2$ 
significantly different from $\bm \beta^1$, leading to a coherent HRS signal for the motion illustrated in Fig. 3c,
while $\bm \alpha ^2 = \bm \alpha ^1$, cancelling the coherent RS response.
A detailed calculation of the coherent enhancement would require a {\em large} simulation, far beyond the scope of this work.
Indeed, reproducing correctly the RS and HRS boson-peak shapes and intensities is likely to be an exacting test of the
simulated structure.

It is finally of interest to compare various BPs as illustrated in the inset of Fig. 2.
The maximum of the HRS BP is at $\Omega {\rm ^{HRS}_{BP}} \simeq 23$ cm$^{-1}$,
while in RS we find $\Omega {\rm ^{RS}_{BP}} \simeq 26$ cm$^{-1}$.
In INS, also performed on a $^{11}$B$_2$O$_3$ sample \cite{Eng98}, an {\em out-of-phase} (incoherent in our terms) 
contribution was separated from an {\em in-phase} one.
The former matches quite well the RS BP, while the latter peaks at a much lower value $\sim$16 cm$^{-1}$.
The sum of these two components, shown in the inset, is positioned at $\Omega {\rm ^{INS}_{BP}} \simeq 19$ cm$^{-1}$.
That coherence lowers $\Omega {\rm_{BP}}$  is consistent with our finding on $\Omega {\rm ^{HRS}_{BP}}$.
However, INS detects in-phase displacements along a line joining neighbors \cite{Eng99}, while the optical
scattering should be more sensitive to rigid rotations.
This could presumably explain the different coherent enhancements leading to $\Omega {\rm ^{HRS}_{BP}} \neq  \Omega {\rm ^{INS}_{BP}}$.
Our analysis here, and in $v$-SiO$_2$ \cite{Heh00}, demontrates that diverse spectroscopies project differently the BP-vibrations, 
leading to significant changes in BP shapes and positions. 
A discussion just in terms of a single light-to-vibration coupling coefficient C($\omega$) \cite{Shu71} 
is presumably too reductive to reflect the complexity of the real situation which involves librations {\em vs.} translations,
coherence {\em vs.} incoherence, and unlike sensitivities of the various spectroscopies to the mode eigenvectors.

The authors thank M.-H. Chopinet and P. Lambremont from Saint-Gobain Recherche for their guidance in the sample preparation,
and P. Umari and A. Pasquarello at EPFL, Lausanne, for confirming by simulation that the boroxol
contribution is the highest frequency one in the 700 cm$^{-1}$ vibrational group \cite{Pas05}.


\begin{thebibliography}{99}
\bibitem{foot} In the molecular-fixed reference frame, $z is \parallel$ to the three-fold axis and
$x is \parallel$ to one of the two-fold axes.
\end{thebibliography}
\end{document}